\documentclass[a4paper]{aa}  
\usepackage{graphicx}
\usepackage{txfonts}
\usepackage{amsmath}
\usepackage[dvipsnames]{xcolor}
\usepackage[colorlinks=true,allcolors=blue]{hyperref}

\newcommand{\bz}{$\langle B_{\rm z} \rangle$}
\newcommand{\bs}{$\langle B \rangle$}
\newcommand{\bd}{$B_{\rm d}$}
\newcommand{\bzrms}{$\langle B_{\rm z} \rangle_{\rm rms}$}
\newcommand{\bsavg}{$\langle B \rangle_{\rm avg}$}
\newcommand{\bzrmsr}{$\langle B_{\rm z} \rangle_{\rm rms} / B_{\rm d}$}
\newcommand{\bsavgr}{$\langle B \rangle_{\rm avg} / B_{\rm d}$}

\newcommand{\figps}[3]{\resizebox{#1}{!}{\rotatebox{#2}{\includegraphics{#3}}}}
\newcommand{\cla}[1]{{#1}}

\begin{document}

\title{Statistical relations between spectropolarimetric observables and the polar strength of the stellar dipolar magnetic field} 

\titlerunning{Estimating dipolar field strength}

\author{
O. Kochukhov
}

\institute{
Department of Physics and Astronomy, Uppsala University, Box 516, S-75120 Uppsala, Sweden\\
\email{oleg.kochukhov@physics.uu.se}
}

\date{Received 2 February 2024 / Accepted 13 April 2024}

\abstract{
Global magnetic fields of early-type stars are commonly characterised by the mean longitudinal magnetic field $\langle B_{\rm z} \rangle$ and the mean field modulus $\langle B \rangle$, derived from the circular polarisation and intensity spectra, respectively. Observational studies often report a root mean square (rms) of $\langle B_{\rm z} \rangle$ and an average value of $\langle B \rangle$. In this work, I used numerical simulations to establish statistical relationships between these cumulative magnetic observables and the polar strength, $B_{\rm d}$, of a dipolar magnetic field. I show that in the limit of \cla{many} measurements randomly distributed in rotational phase, $\langle B_{\rm z} \rangle_{\rm rms}$\,=\,$0.179^{+0.031}_{-0.043}$\,$B_{\rm d}$ and $\langle B \rangle_{\rm avg}$\,=\,$0.691^{+0.020}_{-0.023}$\,$B_{\rm d}$. The same values can be recovered with only three measurements, provided that the observations are distributed uniformly in the rotational phase. These conversion factors \cla{are suitable for} ensemble analyses of large stellar samples, where each target is covered by a small number of magnetic measurements.
}

\keywords{techniques: polarimetric -- stars:  early-type -- stars: magnetic field}

\maketitle

\section{Introduction}

A fraction of early-type, main sequence stars, with spectral types ranging from B to F, exhibit stable and globally organised magnetic fields on their surfaces \citep{donati:2009,wade:2016,sikora:2019a}. These fields, ranging in strength from a few hundred G to tens of kG, give rise to inhomogeneous surface chemical distributions \citep[e.g.][and references therein]{kochukhov:2017}, vertical chemical stratification \citep[e.g.][]{leblanc:2009}, and radio-emitting magnetospheres \citep{das:2022}. As magnetic stars rotate, their non-uniform surfaces are seen from varying aspect angles by a remote observer, resulting in periodic changes of brightness, broad-band photometric colours, spectral energy distributions, and line profiles.

Two types of direct magnetic field measurements are most commonly applied to detect and characterise magnetic fields of early-type stars. We can analyse Zeeman-induced circular polarisation in spectral lines using photopolarimetric methods \citep{landstreet:1980,bohlender:1993}, low- \citep{bagnulo:2002a,bagnulo:2015}, \cla{medium- \citep{monin:2012,semenko:2022},} and high-resolution \citep{mathys:1997a,wade:2000} spectropolarimetry, deriving the so-called mean longitudinal magnetic field, \bz,\ \citep{mathys:1991}. This magnetic observable corresponds to the weighted average of the line-of-sight component of the stellar magnetic field over the stellar disk. On the other hand, it is also possible to obtain the disk-average absolute value of magnetic field, the mean field modulus, \bs, by measuring separation of the Zeeman-split line components in the optical \citep{mathys:1997b,mathys:2017} and near-infrared \citep{chojnowski:2019} high-resolution spectra. In some studies, \bs\ was deduced with the help of detailed theoretical line profile modelling, even when no resolved Zeeman components were detectable in stellar spectra \citep[e.g.][]{kochukhov:2004e,kochukhov:2006b,kochukhov:2013c}.

Both the mean longitudinal magnetic field, \bz,\ and the mean field modulus, \bs,\ vary periodically with stellar rotation. The rotational phase curves of \bz\ and \bs\ typically exhibit a smooth single- or double-wave behaviour. This suggests that the global magnetic topologies of early-type stars are dominated by dipolar components. With  a few exceptions
notwithstanding \citep{donati:2006b,kochukhov:2011a}, this conclusion has been reinforced by Zeeman Doppler imaging \citep[ZDI,][]{kochukhov:2016} studies, where dipolar-like magnetic configurations were retrieved as a result of Stokes profile inversions carried out without any prior assumptions on the global field geometries \citep[e.g.][]{kochukhov:2014,kochukhov:2017a,kochukhov:2019,kochukhov:2023a}. These observational results, along with theoretical modelling of equilibrium fossil magnetic fields in radiative stellar interiors \citep{braithwaite:2006,duez:2010}, validate the usefulness of dipolar field as a first-order approximation of the surface magnetic field topology of early-type stars.

The \cla{most common} method for deriving the parameters of dipolar magnetic field, such as the polar strength \bd, involves collecting a large number of \bz\ and, for a smaller number of targets, \bs\ measurements; then  the resulting phase curves are fitted with dipolar models \citep[e.g.][]{landstreet:2000,auriere:2007,bagnulo:2002,sikora:2019a,shultz:2019a}. This is a time-consuming methodology that requires significant investments of observing time and a prior knowledge of the stellar rotational period. In the present paper, I explore an alternative possibility of constraining \bd\ using a small number of magnetic measurements. I show that cumulative magnetic observables calculated from a few \bz\ or \bs\ observations can be statistically related to \bd, opening up prospects for estimating dipolar field strength for large stellar samples.

\cla{Whenever high-resolution spectropolarimetric data are available, a multitude of diagnostic methods can be applied to extract parameters of global stellar magnetic fields. This includes analyses of higher order moments of Stokes profiles \citep{mathys:1995a}, utilising cumulative integral of Stokes $V$ spectra \citep{kochukhov:2015a,gayley:2017}, applying a principal component analysis \citep{martinez-gonzalez:2008,lehmann:2022}, Bayesian inference \citep{petit:2012}, forward Stokes profile modelling \citep{bagnulo:2001}, and ZDI. These techniques are particularly powerful when applied to Doppler-broadened Stokes profiles of rapidly rotating stars. However, restrictions in terms of stellar parameters and the type of required observational data make these approaches less suitable for large-scale statistical studies compared to traditional \bz\ diagnostic.}

\section{Methods and analysis}

\subsection{Magnetic observables}

In situations when the \bz\ phase curves are incomplete or when considering statistical properties of large stellar samples, a small number of individual longitudinal field measurements cannot be directly used to assess the intrinsic stellar field strength due to a strong rotational phase dependence of \bz. To alleviate this problem, it is customary to convert a set of longitudinal field measurements to a cumulative observable known as the root mean square (rms) longitudinal field \citep{borra:1983,thompson:1987,bohlender:1993}. This quantity, defined as: 
\begin{equation}
\langle B_{\rm z} \rangle_{\rm rms} = \left( \dfrac{1}{N} \sum_{i=1}^N \langle B_{\rm z} \rangle_i^2 \right)^{1/2},
\label{eq:bzrms}
\end{equation}
has been catalogued for over a thousand early-type stars \citep{bychkov:2003,bychkov:2009,hubrig:2006b,romanyuk:2008}, including several samples of stars in young open clusters \citep{semenko:2022,romanyuk:2023}. The field strength statistics based on \bzrms\ has been extensively used to study the origin and evolution of the magnetic fields in the upper main sequence stars \citep{kochukhov:2006,hubrig:2007b,landstreet:2007,landstreet:2008} and to assess the overall distribution function of field strengths \citep{kholtygin:2010,medvedev:2017,makarenko:2021}.

The equivalent cumulative mean magnetic field modulus observable is the average \bs,
\begin{equation}
\langle B \rangle_{\rm avg} = \dfrac{1}{N} \sum_{i=1}^N \langle B \rangle_i.
\label{eq:bsavg}
\end{equation}
Its usage is somewhat less common in the literature, reflecting a considerably smaller number of stars with \bs\ time series compared to repeated \bz\ observations. Nevertheless, \bsavg\ has been compiled by several studies \citep{mathys:1997b,mathys:2017,chojnowski:2019,giarrusso:2022}, resulting in a data base of over 200 stars.

\subsection{Dipolar magnetic field}

Assuming a centred dipolar magnetic field geometry with a polar strength, \bd\, and a linear limb-darkening law \cla{specifying variation of the continuum intensity, $I,$ as a function of the cosine, $\mu,$ of the limb angle,}
\begin{equation}
\cla{
I(\mu) = 1 - u + u \mu,
}
\end{equation}
we can derive the following analytical relations for the rotational phase curves of \bz\ and \bs\ \citep{hensberge:1977,leroy:1994}:
\begin{equation}
\langle B_{\rm z} \rangle = B_{\rm d} C_1(u) \cos{\gamma}
\label{eq3}
\end{equation}
and
\begin{equation}
\langle B \rangle = B_{\rm d} \left[ C_2(u)\cos^2\gamma + C_3(u)\sin^2\gamma \right].
\label{eq4}
\end{equation}
Here, the parameters $C_1$--$C_3$ are functions of the linear limb-darkening coefficient, $u$,
\begin{align}
C_1 (u)  &=  \dfrac{15 + u}{20(3-u)}, \nonumber \\
C_2 (u)  &=  \dfrac{3}{3-u} \left(0.77778 - 0.22613 u\right), \\
C_3 (u)  &=  \dfrac{3}{3-u} \left(0.64775 - 0.23349 u\right), \nonumber
\end{align}
and $\gamma$ corresponds to the angle between the dipolar field axis and the line of sight. This angle can be calculated, for a spherical stellar surface, from the stellar inclination angle, $i$, the magnetic obliquity angle, $\beta$, and the phase angle, $\varphi$ as follows:
\begin{equation}
\cos{\gamma} = \cos{i} \cos{\beta} + \sin{i} \sin{\beta} \cos{\varphi}.
\end{equation}
In this equation, the angles $i$ and $\beta$ can take values in the interval $[0,\pi]$ and are fixed for a given star, whereas $\varphi$ varies between 0 and $2\pi$ in the course of stellar rotation.

\subsection{Numerical simulations}
\label{sec:calc}

In the present paper, I use numerical simulations to establish statistical relations between \bzrms\ and \bsavg\  on the one hand and the dipolar field strength, \bd, on the other hand. In these calculations, I employed Eqs.~(\ref{eq3}) and (\ref{eq4}) and postulated an isotropic distribution of the stellar rotational and magnetic axes. This was numerically implemented by sampling $i$ and $\beta$ according to:
\begin{align}
 i &= \arccos{r_1}, \\
 \beta &= \arccos{r_2}, \nonumber
\end{align}
where $r_1$ and $r_2$ are independent random numbers drawn from a uniform distribution between $-1$ and $+1$. Another set of uniformly distributed random numbers $r_3$ between 0 and 1 was used to assign $N$ rotational phases, $\varphi =2\pi r_3$. I considered two possibilities for phase sampling. In the first case, all $N$ phases were chosen randomly. In the second case, the first phase was selected randomly and the remaining $N-1$ phases were calculated assuming an equidistant phase sampling with a step of $1/N$. These two scenarios correspond to the situation when the stellar rotational period is unknown prior to magnetic observations (the first case) and when this period is known and the timing of observations can be planned accordingly (the second case).

I considered $N$ from 1 to 30, performing calculations for $10^6$ random combinations of $i$, $\beta$ and one of the two options of defining random sets of $\varphi$ angles for each $N$. Calculations were carried out for a single value of the linear limb-darkening coefficient, $u=0.5$. This choice, yielding $C_1=0.310$, $C_2=0.798$, and $C_3=0.637$, roughly corresponds to the $V$-band continuum limb darkening of a main sequence star with solar metallicity and the A0 spectral type \citep{claret:2000,pecaut:2013}. The function $C_1(u)$ varies by about $\pm6$\% around the assumed value for the entire 7000--20\,000~K $T_{\rm eff}$ range, where global magnetic fields are typically found in the upper main sequence stars. Considering the linear dependence of \bzrms\ on $C_1$, we can rescale the results presented below for any desired value of $u$. On the other hand, the functions $C_2(u)$ and $C_3(u)$ change by less than 1\% in the same $T_{\rm eff}$ interval; so their variation with the stellar temperature can be safely neglected.

The resulting probability density functions of the \bzrmsr\ and \bsavgr\ ratios are shown in Fig.~\ref{fig:pdf1} for a random sampling and in Fig.~\ref{fig:pdf2} for an equidistant rotational phase sampling. The median values of these distributions are plotted as a function of $N$ in Fig.~\ref{fig:med} and are reported in Table~\ref{tbl}. The numerical results are presented in this table only up to $N=3$ for the equidistant phase sampling case since there is no change in the shape of distributions for larger $N$ values. Figs.~\ref{fig:pdf1}--\ref{fig:med} and Table~\ref{tbl} also provide confidence intervals containing 68.3\%, 95.5\%, and 99.7\% of the simulation results (i.e. 1-, 2-, and 3-$\sigma$ intervals of a normal distribution). \cla{I note that the $N=1$ calculation for both phase sampling cases corresponds to different realisations of essentially the same numerical test. Accordingly, there is no discernible difference between the top panels of Figs.~\ref{fig:pdf1} and \ref{fig:pdf2}. A small discrepancy in the corresponding numbers in Table~\ref{tbl} reflects numerical uncertainty associated with establishing percentiles for a distribution lacking a well-defined central peak.}

\section{Discussion}

\begin{figure}[!t]
\centering
\figps{\hsize}{0}{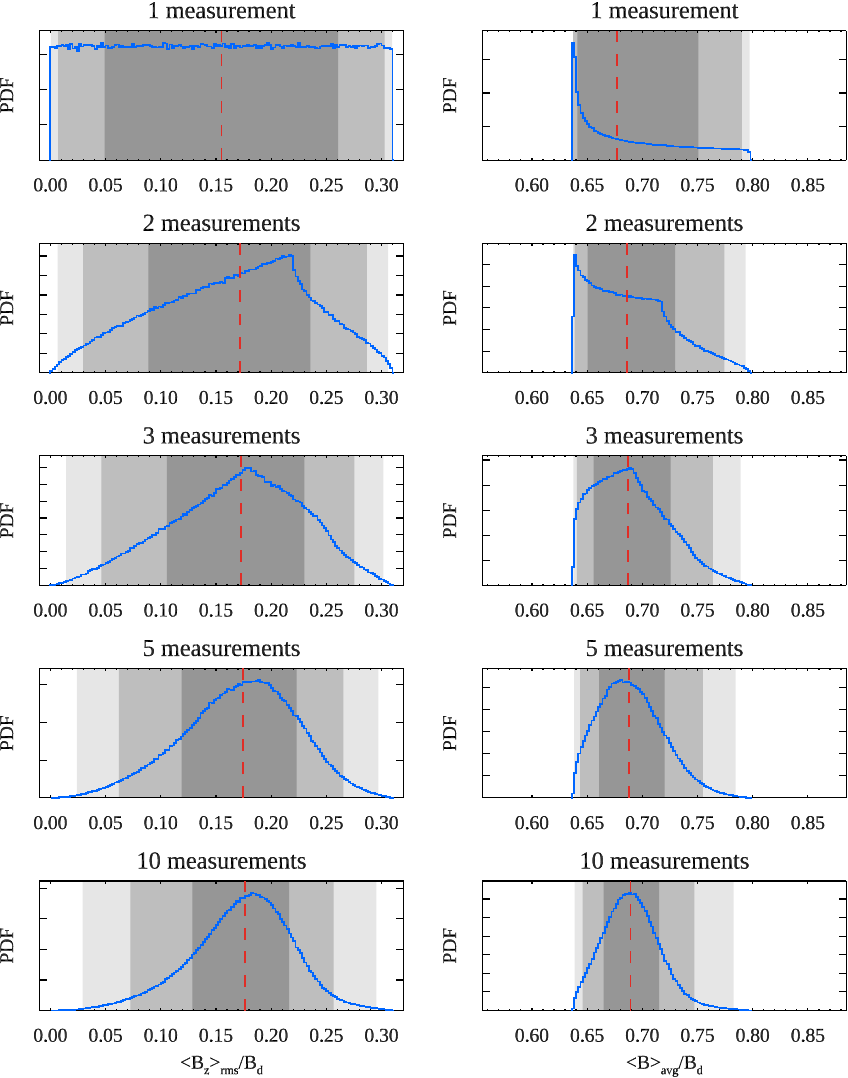}
\caption{Probability distributions of the \bzrmsr\ (left column) and \bsavgr\ (right column) ratios for different number of randomly distributed measurements. The vertical dashed line corresponds to the median of each distribution. The grey rectangles in the background indicate the 1-, 2-, and 3-$\sigma$ confidence intervals.}
\label{fig:pdf1}
\end{figure}

\begin{figure}[!t]
\centering
\figps{\hsize}{0}{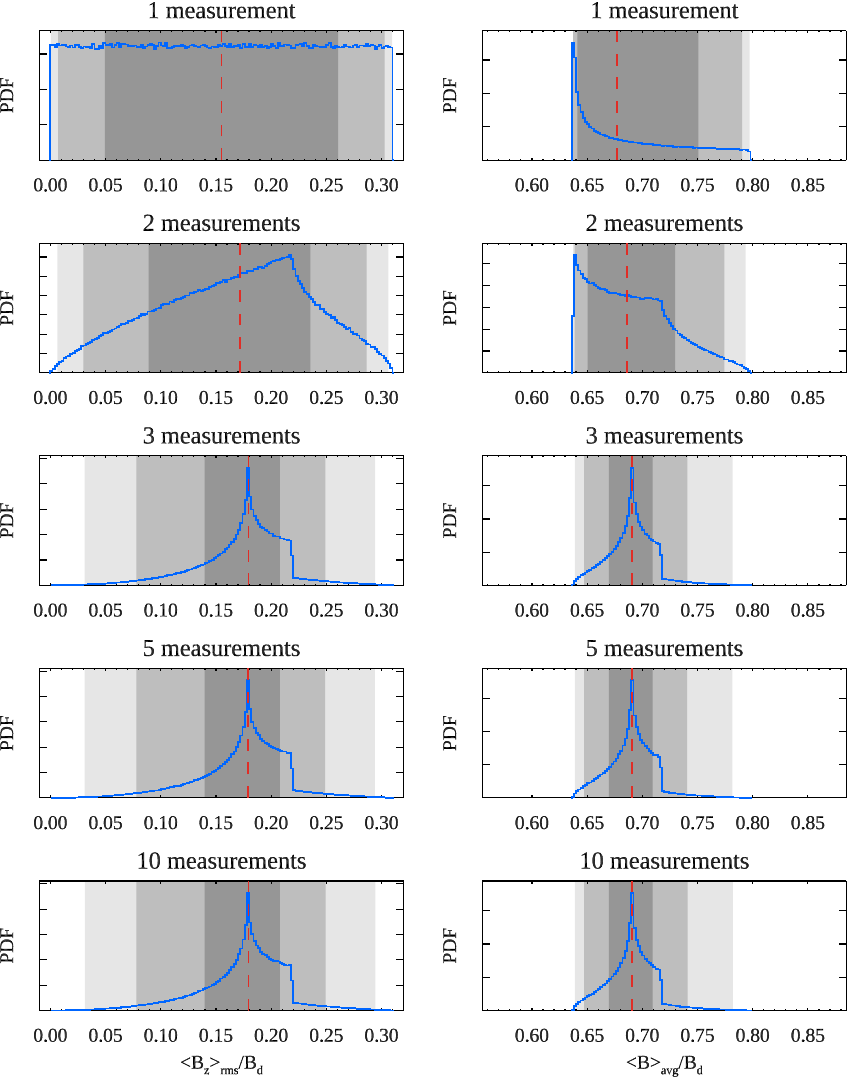}
\caption{Same as Fig.~\ref{fig:pdf1}, but for measurements equidistant in rotational phase.}
\label{fig:pdf2}
\end{figure}

\begin{table*}
\centering
\caption{Median values and confidence intervals for the \bzrmsr\ and \bsavgr\ ratio distributions for different number of magnetic measurements distributed randomly and equidistantly in rotational phase.\label{tbl}}
\begin{tabular}{rcccc|cccc}
\hline\hline
     \multicolumn{9}{c}{Random phase distribution} \\
    & \multicolumn{4}{c|}{\bzrmsr} & \multicolumn{4}{c}{\bsavgr} \\
$N$ & Median & 68.3\% & 95.5\% & 99.7\% & Median & 68.3\% & 95.5\% & 99.7\% \\
\hline
  1 &  0.155 & 0.049--0.261 & 0.007--0.303 & 0.000--0.310 &  0.677 & 0.641--0.751 & 0.637--0.790 & 0.637--0.797 \\
  2 &  0.172 & 0.089--0.235 & 0.030--0.287 & 0.006--0.306 &  0.686 & 0.650--0.730 & 0.639--0.774 & 0.637--0.794 \\
  3 &  0.173 & 0.106--0.230 & 0.046--0.276 & 0.014--0.302 &  0.687 & 0.656--0.726 & 0.641--0.764 & 0.638--0.789 \\
  5 &  0.175 & 0.119--0.223 & 0.062--0.265 & 0.024--0.297 &  0.688 & 0.661--0.720 & 0.644--0.755 & 0.638--0.785 \\
 10 &  0.177 & 0.129--0.216 & 0.072--0.256 & 0.029--0.295 &  0.689 & 0.665--0.715 & 0.646--0.747 & 0.639--0.782 \\
 20 &  0.178 & 0.134--0.212 & 0.076--0.252 & 0.030--0.295 &  0.690 & 0.667--0.712 & 0.647--0.743 & 0.639--0.782 \\
 30 &  0.179 & 0.136--0.210 & 0.076--0.251 & 0.030--0.295 &  0.691 & 0.668--0.711 & 0.647--0.742 & 0.639--0.782 \\
\hline
     \multicolumn{9}{c}{Equidistant phase distribution} \\
    & \multicolumn{4}{c|}{\bzrmsr} & \multicolumn{4}{c}{\bsavgr} \\
$N$ & Median & 68.3\% & 95.5\% & 99.7\% & Median & 68.3\% & 95.5\% & 99.7\% \\
\hline
  1 &  0.149 & 0.051--0.254 & 0.007--0.302 & 0.000--0.309 &  0.674 & 0.642--0.745 & 0.637--0.789 & 0.637--0.797 \\
  2 &  0.172 & 0.089--0.236 & 0.030--0.287 & 0.006--0.306 &  0.686 & 0.650--0.730 & 0.639--0.775 & 0.637--0.794 \\
  3 &  0.179 & 0.140--0.208 & 0.078--0.249 & 0.031--0.294 &  0.691 & 0.670--0.709 & 0.647--0.741 & 0.639--0.782 \\
\hline
\end{tabular}
\end{table*}

The results presented in the upper panels of Fig.~\ref{fig:pdf1} or Fig. \ref{fig:pdf2} demonstrate that a single measurement of either \bz\ or \bs\ is difficult to relate to \bd. The distribution of \bzrmsr\ is flat between 0 and 0.31, indicating that a more appropriate interpretation of a single \bz\ data point is calculating a lower limit \bd\,$\ge$\,$|\langle B_{\rm z} \rangle|/C_1$ or \bd\,$\ge$\,$3.226|\langle B_{\rm z} \rangle|$ for the present choice of the limb-darkening coefficient $u=0.5$. Similarly, a single \bs\ measurement can be interpreted in terms of the upper and lower limits of the dipolar field strength, 1.254\,\bs\,$\le$\,\bd\,$\le$\,1.569\,\bs.

By the time the three measurements at random rotational phases become available, the \bzrmsr\ and \bsavgr\ PDFs start resembling a unimodal, normal-like distribution. In fact, \bzrmsr\ changes by merely 3\% and \bsavgr\ by less than 1\% when going from $N=3$ to $N=30$. At the same time, the 1-$\sigma$ confidence intervals shrink by about 40\% for both observables, illustrating the benefit of $N>3$ measurements. In the limit of a large number of randomly distributed data points, \bzrms\,=\,$0.179\substack{+0.031 \\ -0.043}$\,\bd\ and \bsavg\,=\,$0.691^{+0.020}_{-0.023}$\,\bd, where the quoted uncertainties correspond to the 1-$\sigma$ confidence intervals. Inverting these numbers yields conversion factors $B_{\rm d}/\langle B_{\rm z} \rangle_{\rm rms}$\,=\,$5.59\substack{+1.77 \\ -0.82}$ and $B_{\rm d}/\langle B \rangle_{\rm avg}$\,=\,$1.447\substack{+0.050 \\ -0.041}$, which can be used in the context of large surveys and statistical stellar magnetism studies. In all cases, \bs\ observations provide a tighter constraint on \bd\ than the same number of \bz\ measurements.

\begin{figure}
\centering
\figps{0.49\hsize}{0}{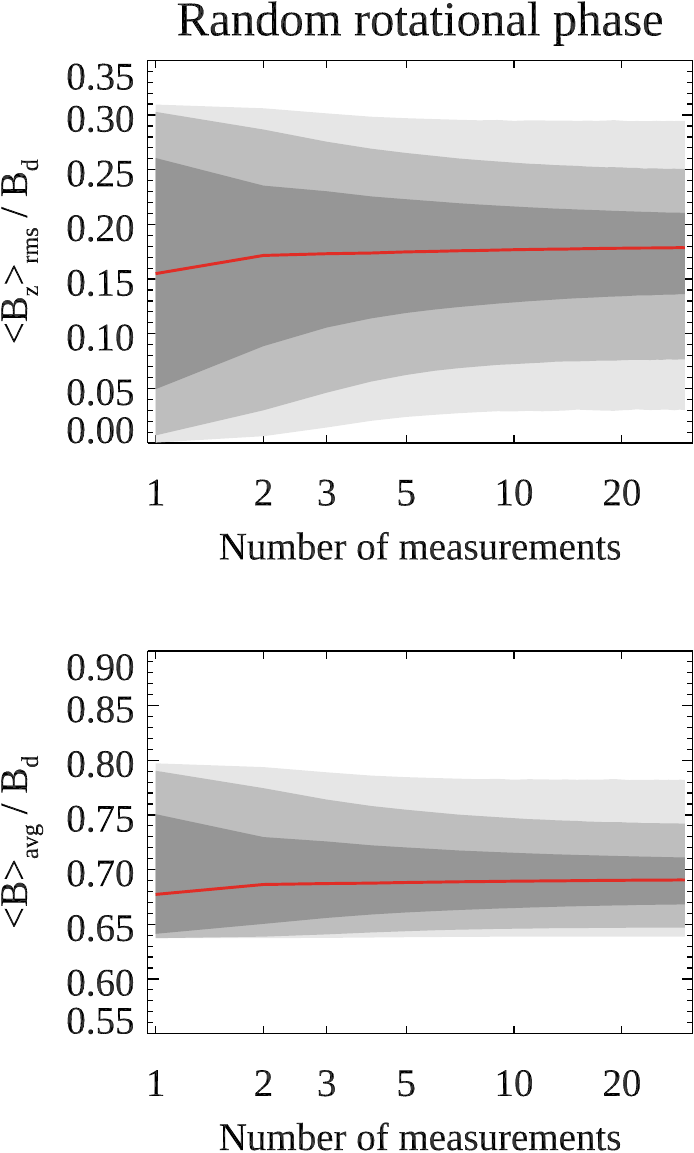}
\figps{0.49\hsize}{0}{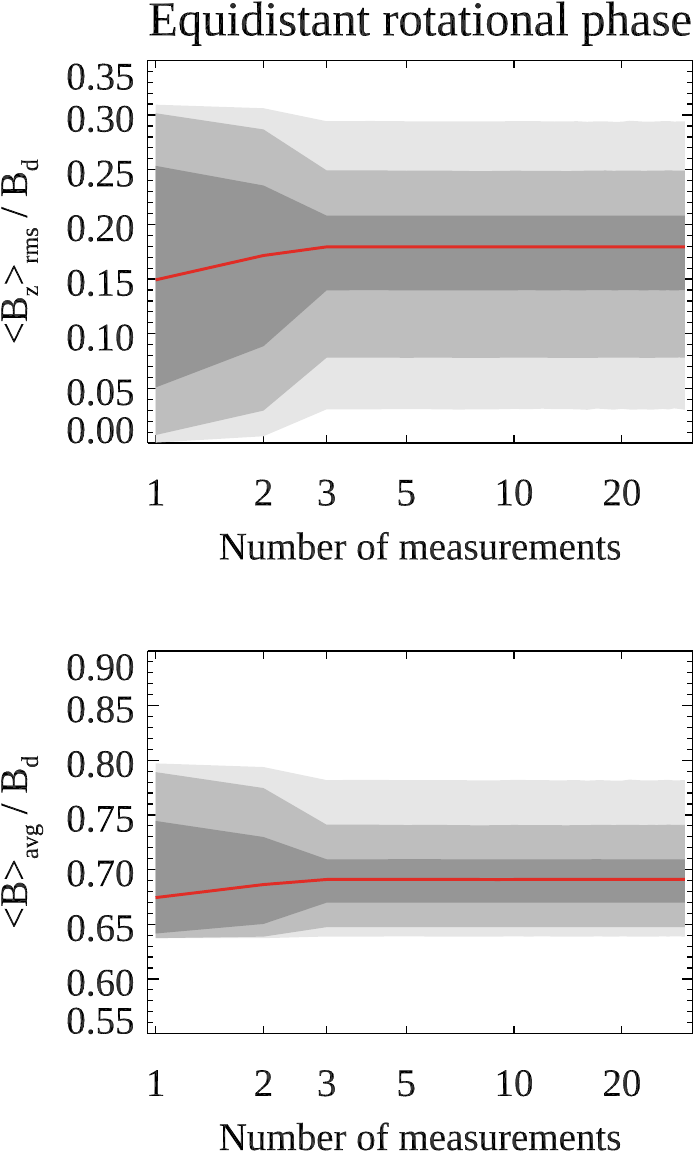}
\caption{Median values (solid red line) and the 1-, 2-, and 3-$\sigma$ confidence intervals (grey outlines) of the \bzrmsr\ (top row) and \bsavgr\ (bottom row) ratios as a function of the number of magnetic measurements distributed randomly (left column) and equidistantly (right column) in the rotational phase.}
\label{fig:med}
\end{figure}

It is interesting to note that in the case of equidistantly spaced observations, the asymptotic \bzrmsr\ and \bsavgr\ ratios determined above are already recovered  at $N=3$ and do not change with increasing number of measurements. The corresponding confidence intervals do not improve either (see Fig.~\ref{fig:med}). In this case, the PDFs settle on unimodal, but distinctly asymmetric and non-Gaussian probability distributions. Another way to arrive at these distributions is to replace the sums in Eqs.~(\ref{eq:bzrms}) and (\ref{eq:bsavg}) with an integral over $\varphi$, yielding:
\begin{align}
\langle B_{\rm z} \rangle_{\rm rms} &= B_{\rm d} C_1(u) \sqrt{\langle \cos^2{\gamma} \rangle }, \\
\langle B \rangle_{\rm avg} &= B_{\rm d} \left[ C_2(u) \langle \cos^2{\gamma} \rangle + C_3(u) \langle \sin^2{\gamma} \rangle \right],
\end{align}
with the phase-averaged $\cos^2\gamma$ and $\sin^2\gamma$ functions given by
\begin{align}
\langle \cos^2{\gamma} \rangle & = \dfrac{1}{8} \left[ 3 + \cos{2i} + \cos{2\beta} (1 + 3 \cos{2i}) \right], \\
\langle \sin^2{\gamma} \rangle & = \dfrac{1}{8} \left[ 5 - \cos{2i} - \cos{2\beta} (1 + 3 \cos{2i}) \right],
\end{align}
and the angles $i$, $\beta$ sampled with isotropic distributions as before.

For completeness, I also calculated the ratio of the rms longitudinal field to the average field modulus, $\langle B_{\rm z} \rangle_{\rm rms} / \langle B \rangle_{\rm avg}$\,=\,0.259$\substack{+0.037\\-0.055}$, valid for $N=30$ in the first phase sampling case and $N\ge3$ in the second one. This corresponds to the conversion \bsavg\,=\,3.86$\substack{+1.04\\-0.48}$\bzrms, which could be used to obtain from \bzrms\ a representative average field strength parameter rather than its extreme value at the magnetic poles.

The importance of spreading out a few magnetic measurements over stellar rotational cycle is a noteworthy conclusion of the present study. The perfectly equidistant sampling considered for the calculations in Sect.~\ref{sec:calc} is evidently an idealisation. Nevertheless, it is not too distant from reality since the information on rotational periods of early-type stars is  readily available at present from decades of ground-based photometric observations \citep[e.g.][]{hummerich:2016,netopil:2017,bernhard:2020} and from the high-precision spaceborne light curves \citep{wraight:2012,cunha:2019,holdsworth:2021,holdsworth:2024}. It is therefore feasible to carefully plan magnetic observations of stars with known rotational properties, thereby minimising the number of required data points.

\begin{acknowledgements}
I acknowledge support by the Swedish Research Council (grant agreements no. 2019-03548 and 2023-03667), the Swedish National Space Agency, and the Royal Swedish Academy of Sciences.
This research was also partly supported by the Munich Institute for Astro-, Particle and BioPhysics (MIAPbP) which is funded by the Deutsche Forschungsgemeinschaft (DFG, German Research Foundation) under Germany's Excellence Strategy -- EXC-2094 -- 390783311.
\end{acknowledgements}


\end{document}